\documentclass[aps,pre,twocolumn,amsmath]{revtex4}
\usepackage{graphicx}
\usepackage{bm}
\usepackage{epstopdf}
\usepackage{dcolumn}

\begin{document}

\title{Hopping Transport in Hostile Reaction-Diffusion Systems}
\author{Andrew R. Missel}
\email[]{missel@uiuc.edu}
\author{Karin A. Dahmen}
\email[]{dahmen@uiuc.edu}
\affiliation{Physics Department, University of Illinois at Urbana-Champaign, Urbana, IL 61801, USA}

\date{\today}

\begin{abstract}
We investigate transport in a disordered reaction-diffusion (RD) model consisting of particles which are allowed to diffuse, compete with one another ($2A\to A$), give birth in small areas called ``oases'' ($A\to2A$), and die in the ``desert'' outside the oases ($A\to0$).  This model has previously been used to study bacterial populations in the lab and is related to a model of plankton populations in the oceans.  We first consider the nature of transport between two oases: in the limit of high growth rate, this is effectively a first passage process, and we are able to determine the first passage time probability density function in the limit of large oasis separation.  This result is then used along with the theory of hopping conduction in doped semiconductors to estimate the time taken by a population to cross a large system.
\end{abstract}

\maketitle

\section{Introduction}
\subsection{Reaction Diffusion Models}
Reaction-diffusion (RD) models have proven to be very useful tools for the study of chemical \cite{kroon1993}, biological \cite{murray}, and ecological \cite{birch2007bbf} systems.  RD models typically consist of a set of particles which are allowed to diffuse and interact with one another and their environment in prescribed ways.  By varying the types of allowed reactions, number of types of particles, and reaction rates, one can obtain a wide variety of behavior.  Much work has been done to examine the phase transition between active (population survives as $t\to\infty$) and absorbing (population dies as $t\to\infty$) states \cite{grassberger1979rft,janssen1981zpb,hinrichsenreview,janssentauberreview, tauberftrev} and to determine the nature of propagating fronts \cite{panjareview,debierre1994fpr,dickmaninterface,moro2001ife}.

Typically, RD models are governed by a microscopic master equation \cite{vankampenbook} which describes the probability flow into and out of the microstates of the system.  This master equation is not solvable for all but the most simple models, and thus various approximation techniques---Langevin equations, for example---are usually used.  There does exist a systematic expansion of the master equation \cite{vankampenbook}, the lowest order of which is usually a deterministic differential equation or Fokker-Planck equation for the mean concentrations of the constituent particles.  These equations---reaction-diffusion equations---are often studied first as a means of characterizing the qualitative behavior of the model under examination; they constitute a mean-field theory for the model.

The effects of quenched disorder in the reaction rates on the critical behavior of RD models have been difficult to determine.  A straightforward renormalization group treatment leads to runaway flows \cite{tauberftrev,janssen1997rft}, but some progress has been made using simulations \cite{MCofcontactproc, webman1998dba,szaboPRE2002} and real-space RG methods \cite{hooyberghs2003sdf, hooyberghs2004asp}.  Disorder effects on RD fronts have also been studied, mostly for the case in which the disorder is time-dependent (``annealed'') and the system admits a front solution in the absence of noise \cite{lemarchand1995lac,grinsteinprl96,flucsinfronts, flucsinfronts2} .  However, a few studies have been made of the effects of quenched disorder on RD fronts \cite{fisherwavesrandommedia}, and some attention has been devoted to the interesting case of noise-induced fronts \cite{noisefronts,sagues2007soo}.

\subsection{Our Model: Oases and Deserts}
This work concerns the nature of transport in a particular reaction-diffusion system with spatial inhomogeneity in the reaction rates.  We study a model with a mean-field limit defined by the generalized Fisher/KPP equation
\begin{equation}
\frac{\partial c(\bm{x},t)}{\partial t}=D\nabla^{2}c(\bm{x},t)-\bm{v}\cdot\nabla c(\bm{x},t)+U(\bm{x})c(\bm{x},t)-qc(\bm{x},t)^{2},\label{FisherKPP}
\end{equation}
where $c(\bm{x},t)$ represents the population density, $D$ is the diffusion constant, $\bm{v}$ is a spatially uniform convection velocity (representing the flow of some liquid in which the particles exist), $U(\bm{x})$ is a spatially inhomogeneous growth term fixed in time, and $q=b{\ell_{0}}^{d}$ is a competition term ($b$ is a competition rate and $\ell_{0}$ is the microscopic length scale at which two particles will compete with one another).  One of the simplest cases to consider is when $U(\bm{x})=-z$ everywhere except a small patch near the origin, where $U(\bm{x})=y$.  The region of positive growth rate near the origin is called an ``oasis,'' while the rest of space is termed the ``desert.''  This model was previously studied by Nelson and coworkers \cite{dahmennelshn1999,dahmennelshn2000}, and a microscopic model (the contact process with disorder) with this mean-field limit was studied by Joo and Lebowitz \cite{joo2005pds}.  Both sets of researchers found a transition in the $\langle U(\bm{x})\rangle$-$|\bm{v}|$ plane between extinct, localized, and delocalized phases in finite systems with periodic boundary conditions: for high average growth rate and high convection velocity, they observed a delocalized phase; for low average growth rate and high convection velocity they found that the population became extinct; and for low average growth rate and low convection velocity they found a localized phase.  These predictions were tested in a laboratory setting using bacteria protected from harmful UV light (the ``desert'') by a mask (the ``oasis''); the experiments largely confirmed the theoretical predictions summarized above \cite{exper2}.

In this paper, we will examine the nature of transport in a system consisting of many identical oases distributed randomly at low density in a desert.  We term this low oasis density regime ``hostile''; the opposite case in which oases fill up most of space we call ``fertile.''  Because transport between oases in such a system involves the movement of a low population density, fluctuations about the mean-field theory (discreteness effects) will be important.  We will thus be examining a particular stochastic process with a mean-field limit given by (\ref{FisherKPP}).  This process is easiest to introduce on a $d=1$ lattice; the generalization to higher dimensions is trivial.  Identical particles (labeled $A$) occupy lattice sites without occupation number limits and are allowed to undergo the following processes: hopping to either side with rate $w/2$ (total hopping rate of $w$); death ($A\to0$) with rate $z$ if in the desert; reproduction ($A\to2A$) with rate $y$ if on an oasis; and competition/coagulation ($2A\to A$) with rate $b$ everywhere.  This process is governed by a master equation for the joint probability $P(\{c\},t)$ to have occupation numbers $\{c\}\equiv\{\ldots,c_{\nu-1},c_{\nu},c_{\nu+1}\ldots\}$ on the lattice points $\nu$ at time $t$:
\begin{widetext}
\begin{eqnarray}
\frac{\partial P(\{c\},t)}{\partial t}&=&\frac{w}{2}\Big(\sum_{\nu}(c_{\nu-1}+1)P(\ldots,c_{\nu-1}+1,c_{\nu}-1,\ldots,t) + (c_{\nu+1}+1)P(\ldots,c_{\nu}-1,c_{\nu+1}+1,\ldots,t)-2c_{\nu}P(\{c\},t)\Big)\nonumber\\
& + & \sum_{\nu}z_{\nu}\left[(c_{\nu}+1)P(\ldots,c_{\nu}+1,\ldots,t)-c_{\nu}P(\{c\},t)\right] + \sum_{\nu}y_{\nu}\left[(c_{\nu}-1)P(\ldots,c_{\nu}-1,\ldots,t)-c_{\nu}P(\{c\},t)\right]\nonumber\\
& + & b\sum_{\nu}(c_{\nu}+1)c_{\nu}P(\ldots,c_{\nu}+1,\ldots,t)-c_{\nu}(c_{\nu}-1)P(\{c\},t).\label{mastereq}
\end{eqnarray}
\end{widetext}
Here $z_{\nu}=0$ on the oases and $z$ in the desert, and $y_{\nu}=0$ in the desert and $y$ on the oases.

Let us now present a brief outline of this paper: in section \ref{1Oasis}, we will examine the nature of growth near a single oasis.  Because the mean-field equation for the steady-state population density is exactly solvable in one dimension, we will be able to identify a length scale describing the distance away from the oasis at which fluctuations about the mean-field theory become important.  We will also briefly discuss in this section the problem of extinction.  In section \ref{2Oases}, we will look at transport between two oases.  By using the fact that the $2A\to A$ competition process is unimportant far away from an oasis where the population is low, we will be able to devise a simpler model which captures the transport characteristics of the full model for large oasis separation.  In section \ref{ManyOases}, we will finally tackle the problem of transport in a system with many oases.  By employing an analogy with the problem of hopping conduction in doped semiconductors, we will estimate the time taken for a population to cross a large system.  Finally, we offer a summary of our results along with some remarks in section \ref{concl}.  Much of the material in sections (\ref{2Oases}) and (\ref{ManyOases}) has been described by us in an earlier publication in less detail \cite{ourPRL-hopping}.  

\section{\label{1Oasis}Growth Near One Oasis}
\subsection{\label{MF1Oasis}Mean-Field Description}
We begin with a study of the nature of population growth near a single oasis in mean-field theory, starting with a $1$D lattice with a single oasis of width $2a$ lattice points centered at the origin.  First, we multiply $P(\{c\},t)$ by $c_{\nu}$ in Eq. (\ref{mastereq}) and sum over configurations to obtain an equation for the time evolution of the average particle concentration $\langle c_{\nu}\rangle(t)$:
\begin{eqnarray}
\frac{\partial\langle c_{\nu}\rangle(t)}{\partial t} & = & \frac{w}{2}\left[\langle c_{\nu+1}\rangle(t)+\langle c_{\nu-1}\rangle(t)-2\langle c_{\nu}\rangle(t)\right]\nonumber\\
& + & \left[y_{\nu}-z_{\nu}\right]\langle c_{\nu}\rangle(t)-b\langle c_{\nu}(c_{\nu}-1)\rangle(t).
\end{eqnarray}
In order to obtain a ``mean-field'' description of our system, we replace the term $\langle c_{\nu}(c_{\nu}-1)\rangle$ with $\langle c_{\nu}\rangle^2$.  This replacement should work well when the population is large---i.e., near the oasis---since we would expect the relative fluctuations in particle number to be smaller in this case.  (There are, of course, more formal ways of deriving the mean-field equation from the master equation.  See, for instance, Ref. \cite{vankampenbook}.)  With this replacement, we can write a mean-field equation for $\bar{c}(\nu,t)\equiv\langle c_{\nu}\rangle(t)$:
\begin{eqnarray}
\frac{\partial \bar{c}(\nu,t)}{\partial t} & = & \frac{w}{2}\left[\bar{c}(\nu+1,t)+\bar{c}(\nu-1,t)-2\bar{c}(\nu,t)\right]\nonumber\\
& + & \left[y(\nu)-z(\nu)\right]\bar{c}(\nu,t)-b\bar{c}(\nu,t)^2.
\end{eqnarray}
It is easier to consider the continuum version of this equation, which is obtained by introducing a lattice spacing $\ell_{0}$ and redefining $\bar{c}(\nu,t)\to\bar{c}(\nu,t)\ell_{0}$, $b\to q/\ell_{0}$, and $\nu\to x/\ell_{0}$.  The diffusion constant $D$ is defined as $w{\ell_{0}}^{2}/2$.  This leads to the $d=1$ version of (\ref{FisherKPP}), with $U(x)=(y+z)\Theta(a-|x|)-z$, where $\Theta(\cdot)$ is the Heaviside step function.  The length scale $\ell_{0}$ has an interpretation in the continuum as the distance within which particles compete with one another.

There are two things we would like to know: first, what does the mean-field concentration $\bar{c}(x,t)$ look like as $t\to\infty$?  Second, what is the time scale on which a small population grows into a substantial population?  Solving analytically for $\bar{c}(x,t)$ for all times is not feasible, but it is possible to solve for the steady-state $t\to\infty$ solution $\bar{c}(x,t=\infty)\equiv\bar{c}_{ss}(x)$ and thus answer the first question.  This function is given by
\begin{displaymath}
\bar{c}_{ss}(x)=\bar{c}_{ss}(0)-m_{+}\,\text{sn}^{2}\left(\sqrt\frac{q|m_{-}|}{6D}\,|x|,\,\imath\sqrt\frac{m_{+}}{|m_{-}|}\,\right)\quad |x|<a
\end{displaymath}
\begin{equation}
\bar{c}_{ss}(x)=\frac{3z}{2q}\text{csch}^{2}\left(\frac{\kappa}{2}(|x|-a)+C\right)\quad |x|>a,\label{c1Dssdes}
\end{equation}
where $\text{sn}(u,k)$ is a Jacobi elliptic function, $\kappa\equiv\sqrt{z/D}$, $\bar{c}_{ss}(0)$ is the steady-state population at the origin, $C=\text{csch}^{-1}(\sqrt{2q\bar{c}_{ss}(a)/3z})$ ($\bar{c}_{ss}(a)$ is the steady-state population at the edge of the oasis), and $m_{+,-}$ are defined as  $\frac{1}{2}\left[3\bar{c}_{ss}(0)-3y/2q\pm\sqrt{\left(3y/2q-\bar{c}_{ss}(0)\right)\left(3y/2q+3\bar{c}_{ss}(0)\right)}\right]$.  The constants $\bar{c}_{ss}(0)$ and $\bar{c}_{ss}(a)$ can be found by matching the solutions and their derivatives at $|x|=a$.  This leads to a transcendental equation for $\bar{c}_{ss}(0)$:
\begin{equation}
\bar{c}_{ss}(0)=\frac{m_{+}\,\text{sn}^{2}\left(\sqrt\frac{q|m_{-}|}{6D}\,a,\,\imath\sqrt\frac{m_{+}}{|m_{-}|}\,\right)}{\left[1-\sqrt{\frac{3y-2q\bar{c}_{ss}(0)}{3(y+z)}}\right]}.\label{c01D}
\end{equation}
Numerically, we have found that an excellent approximation to $\bar{c}_{ss}(0)$ is $\bar{c}_{ss}(0)\simeq(y-y_{c})/q$, where $y_{c}$ is the minimum growth rate at which the population does not die off as $t\to\infty$ when $q=0$.  This cutoff can be found by solving (\ref{FisherKPP}) with $q=0$ (see Appendix \ref{yc}), which leads to the following transcendental equation for $y_{c}$:
\begin{equation}
y_{c}=z\cot^{2}\left(\sqrt{\frac{y_{c}}{D}}\,a\right).\label{ycutoff1D}
\end{equation}

At large distances from the oasis ($|x|\gg a$), $\bar{c}_{ss}(x)\simeq\bar{c}_{\infty}e^{-\kappa|x|}$, where $\bar{c}_{\infty}=4\gamma^{2}\bar{c}_{ss}(a)e^{\kappa a}$ ($\gamma^{-1}=1+\text{csch}(C)$).  In the limit of high growth rate---$y\to\infty$ with all other rates fixed---$\bar{c}_{ss}(a)\to\infty$ and $\bar{c}_{\infty}\to6ze^{\kappa a}/q$.  For smaller values of $y$, $\bar{c}_{ss}(a)$---and thus $\bar{c}_{\infty}$---can be found by first solving for $\bar{c}_{ss}(0)$ using (\ref{c01D}) and then using the relation (see Appendix \ref{1DMFsol} for derivation) $\bar{c}_{ss}(a)=\bar{c}_{ss}(0)\sqrt{\frac{3y-2q\bar{c}_{ss}(0)}{3(y+z)}}$.

In higher dimensions, we consider a hyperspherical oasis of radius $a$.  It is not possible to solve exactly the $t\to\infty$ nonlinear mean-field equation for $d>1$, but it is easy to ascertain the asymptotic behavior of $\bar{c}_{ss}(\bm{x})$ far away from the oasis.  To do so, we drop the nonlinear term from the mean-field equation (\ref{FisherKPP}) under the assumption that $\bar{c}_{ss}(\bm{x})$ is small far from the oasis.  This leads to the linear equation
\begin{equation}
0=D\nabla^{2}\bar{c}_{ss}(\bm{x})-z\bar{c}_{ss}(\bm{x}),
\end{equation}
which is valid far away from the oasis.  In two dimensions, this is solved by $\bar{c}_{ss}(\bm{x})\simeq\bar{c}_{\infty}K_{0}\left(\kappa r\right)$, where $r=|\bm{x}|$ and $K_{0}$ is a modified Bessel function of the first kind.  In three dimensions, $\bar{c}_{ss}(\bm{x})\simeq\bar{c}_{\infty}e^{-\kappa r}/\kappa r$.  Because finding an exact solution for the entire space (including $r<a$) is no longer possible for $d=2$ or $3$, we cannot write down an analytic expression for the prefactors $\bar{c}_{\infty}$ in front of these asymptotic functional forms.

The question of the time scale on which a small population grows into a substantial population has been addressed by Nelson and coworkers \cite{dahmennelshn1999, dahmennelshn2000}.  They analyzed the eigenvalue spectrum of the linearized ($q=0$) version of (\ref{FisherKPP}) and found that the largest eigenvalue $\Gamma_{0}$ is given by \cite{dahmennelshn2000}
\begin{equation}
\Gamma_{0}=(y+z)f\left(\sqrt{D/a^2(y+z)}\right)-z,
\end{equation}
where $f(x)$ is a monotonically decreasing function of $x$ which goes as $1-\pi^2 x^2/4$ for $x\ll1$ and $1/x^2$ for $x\gg1$  In the limit of large $y$, then, $\Gamma_{0}\simeq y$, and the time scale on which a small population grows up is $\sim1/y$.

\subsection{\label{extinc}Fluctuations and Extinction}
It has been known for some time that fluctuations can drive a system to extinction even when mean-field theory predicts a stable active state.  In the case of a continuous \emph{homogeneous} system with the same reactions as our system---$A\to2A$ with rate $y$, $A\to0$ with rate $z$, and $2A\to A$ with rate $b$---there is an active phase only when $z-y<r_{c}$, where $r_{c}$ depends on dimension but is less than zero for $d=1,2,3$ \cite{tauberftrev}.  Mean-field theory, on the other hand, predicts an active phase for $y>z$; fluctuations drive the critical growth rate up.  The disparity between mean-field and stochastic behavior is even greater in the case of a $d=0$ system: mean-field theory predicts a $t\to\infty$ steady state which is reached for any nonzero initial condition, but solving the master equation leads to the conclusion that, for any $z>0$, the population will eventually become extinct \cite{vankampenbook}.  The mean extinction time in this case can be calculated exactly as a function of $y$, $z$, $b$, and the starting population $n_{0}$, although the resulting expression is cumbersome to work with \cite{vankampenbook}.

For the case of a single oasis in an infinite desert, it seems clear that the population will become extinct as $t\to\infty$ for $d=1,2,3$: the finite oasis cannot compete with the infinite desert, regardless of how high the growth rate $y$ is.  For the problem we will be considering, it is important that the oases not die out too early, and thus we need to know the dependence of the mean extinction time on the various parameters of the problem.  The field-theoretic tools used to analyze systems with translational invariance are hard to apply to this case, as are the various methods (see Ref. \cite{meersonPRE07} for one such method) used to analyze $d=0$ systems.  Nonetheless, we can try to place a lower limit on the extinction time.  To do so, we will return to the lattice case in one dimension; our results will be applicable to the continuum case and to other dimensions.  

Consider the case of a perfectly deadly desert, $z\to\infty$.  This effectively turns our system into a finite system with $2a$ lattice points and absorbing boundaries.  The ``effective'' death rate is of the order of $w$, the hopping rate. Now consider a $d=0$ system with the same birth and competition rates which has a death rate of $w$, the hopping rate in our original system.  Our $d=1$ system will certainly live longer than this system, on average: the number of events needed to extinguish the population completely is much larger.  As mentioned above, the mean extinction time for this $d=0$ system can be calculated explicitly, with the result that $T_{\text{extinct}}\sim e^{cy}$, where $c$ is a constant, for large $y$ \cite{vankampenbook}.  This suggests that the mean extinction time should rise at least exponentially with $y$ in our one oasis problem when $y$ is large.  By choosing a large $y$, then, we can ensure that extinction will not invalidate our results.  From here on, we will assume that the growth rate on the oases is large enough that extinction is unlikely on the transport time scales in question.

\section{\label{2Oases}Transport Between Two Oases}
\subsection{Transport as a First Passage Process}
Our eventual goal is to understand the transport of a population across a system filled with oases at low density.  The first step towards such an understanding is to determine the nature of transport between two oases.  Consider two oases of radius $a$ in $d$ dimensions.  The center of one oasis is located at the origin, and the center of the other oasis is located at position $\bm{R}$.  At $t=0$, the first oasis is populated and the second oasis is empty.  We wish to find the infection time---that is, the time it takes for a population to take hold and reach a significant level on the second oasis.  This time can be roughly broken into two parts: $T_{\text{transit}}$, the time it takes particles from the first oasis to reach the second oasis; and $T_{\text{growth}}$, the time it takes the population to rise to a significant level once the second oasis has been reached.  We will assume that the first particle to reach the second oasis will reproduce and that its offspring will not die out; in other words, we will take $T_{\text{transit}}$ to be the first passage time (FPT) of the process.   This assumption can be satisfied in two ways: the first way is simply to make the growth rate $y$ of the oases very high.  In this case, it is possible to estimate how the survival probability increases as $y$ increases.  Consider again the case of a very deadly desert: if the particle diffuses off the oasis, it is certainly dead; thus, there is an effective death rate of order $D/{\ell_{0}}^2$.  For the case of a very small oasis, then, a toy model of the oasis is a $d=0$ system with death rate of $D/{\ell_{0}}^2$.  For this case, it is known that the survival probability goes like $1-D/y{\ell_{0}}^2$ \cite{vankampenbook}, and thus making $y$ very high assures that the population will take hold and survive.  A second way of satisfying our assumption is to seed the oases with a second species of particles, $B$, which interact with the $A$ particles via the reaction $A+B\to2A$ at a very high rate.  

The time $T_{\text{growth}}$ that it takes the initially small population on the second oasis to grow to a macroscopic size should go roughly like $1/y$ for large $y$, and so choosing a large $y$ should also serve to make $T_{\text{growth}}\ll T_{\text{transit}}$.  For the remainder of the paper, we will assume that $y$ is large enough so that this is the case.  Note that by taking $y$ to be very high, we have done three things: first, we have ensured that a small population which reaches a new oasis grows into a sizable population and does not die out, which allows us to identify the first passage time with the transit time; second, we have made the time for this growth small compared to the transit time; and finally, as mentioned in the previous section, we have ensured that extinction will only occur on a time scale much larger than the one associated with transit.

Consider the case where the two oases are close together: particles from the first oasis diffuse out in a front, its amplitude decaying due to the death term in the desert and competition effects.  However, so long as the second oasis is close enough that the edge of the front is almost certain to possess many particles (the number will vary from realization to realization of the stochastic process), the transit time should simply go as $R$, the oasis separation.  However, once $R$ is well above some length scale we will call $R_{\text{lin}}$, this is no longer true: the front simply does not exist in most realizations of the system, as the number of particles present at this distance from the first oasis is quite small for all times.  In this regime, the second oasis is reached not by a front but by a stray particle (or some stray particles) that manages to make it through the desert; it is essentially a noise-induced growth process.  $R_{\text{lin}}$ can thus be roughly defined as the distance from the oasis at which the large-time average concentration falls to $1/{\ell_{0}}^d$.  We have already analyzed the mean-field equations for the average concentration as $t\to\infty$, and found that, except in $d=1$, there are no closed-form solutions.  In one dimension, setting the mean-field $t\to\infty$ average concentration (\ref{c1Dssdes}) for large $y$ equal to $1/\ell_{0}$ and solving for $R_{\text{lin}}$ leads to
\begin{equation}
R_{\text{lin}}=a+\sqrt{\frac{4D}{z}}\text{csch}^{-1}\left(\sqrt{\frac{2b}{3z}}\right).\label{Rlin1D}
\end{equation}
In the limit of large $z/b$, this simplifies to $R_{\text{lin}}\simeq a+\sqrt{D/z}\ln(6z/b)$, where $b$ is $q/\ell_{0}$.  If $y$ is smaller, the relevant length scale will also be smaller.  We believe that this length scale should be of the same order of magnitude in higher dimensions, and so (\ref{Rlin1D}) should also provide a rough estimate of $R_{\text{lin}}$ for $d=2$ and $d=3$.

\subsection{A Simpler Linear Model With a Source}
As we move further from the first oasis, the competition process $2A\to A$ becomes less and less important, especially if $b$ is small compared to the other rates in the problem.  Due to this fact, it is natural to wonder if ignoring these interactions altogether might be the first step in the creation of a tractable model with the same large distance first passage properties as the full model with competition.  We will now propose such a model, which has been discussed by us in an earlier work \cite{ourPRL-hopping}: consider replacing the first oasis with desert, and then placing a point source in the middle that produces non-interacting particles at some average rate $g$.  The master equation for this process on a lattice in $d=1$ can be written as
\begin{widetext}
\begin{eqnarray}\label{linME}
\frac{\partial P(\{n\},t)}{\partial t}&=&\frac{w}{2}\Big(\sum_{\nu}(n_{\nu-1}+1)P(\ldots,n_{\nu-1}+1,n_{\nu}-1,\ldots,t) + (n_{\nu+1}+1)P(\ldots,n_{\nu}-1,n_{\nu+1}+1,\ldots,t)-2n_{\nu}P(\{n\},t)\Big)\nonumber\\
& + & z\sum_{\nu}\left[(n_{\nu}+1)P(\ldots,n_{\nu}+1,\ldots,t)-n_{\nu}P(\{n\},t)\right] + g\left[P(\ldots,n_{0}-1,\ldots,t)-P(\{n\},t)\right],
\end{eqnarray}
\end{widetext}
where $P(\{n\},t)$ is the joint probability to have occupation numbers $\{n\}\equiv\{\ldots,n_{\nu-1},n_{\nu},n_{\nu+1}\ldots\}$ on the lattice points $\nu$ at time $t$.  For an appropriately chosen $g$, the mean flux of particles past the surface at $R_{\text{lin}}$ should match that of the model with competitions; beyond that point, the model with a source differs from the model with competitions only in that it ignores the rare annihilation interactions between particles.  We will show that, for an appropriately chosen $g$, this model---which we will refer to as the linear model with a source---accurately captures the first passage properties of the full nonlinear model with competition.

As with the full nonlinear model with competition (hereafter referred to as the nonlinear model), it is useful to analyze the mean-field behavior of the linear model with a source.  The master equation (\ref{linME}) can be multiplied by $n_{\nu}$ and summed over configurations to obtain an equation for the time evolution of the average number of particles $\bar{n}(\nu,t)$:
\begin{eqnarray}
\frac{\partial\bar{n}(\nu,t)}{\partial t} & = & \frac{w}{2}\left[\bar{n}(\nu+1,t)+\bar{n}(\nu-1,t)-2\bar{n}(\nu,t)\right]\nonumber\\
& & -z\bar{n}(\nu,t)+g\delta_{\nu,0}.\label{MFlinwsource}
\end{eqnarray}
We will study the continuum version of this equation in detail in one, two, and three dimensions.  Taking the continuum limit of (\ref{MFlinwsource}) (and changing $\partial_{x}^{2}\to\nabla^{2}$ for $d>1$) results in:
\begin{equation}
\frac{\partial\bar{n}(\bm{x},t)}{\partial t}=D\nabla^{2}\bar{n}(\bm{x},t)-z\bar{n}(\bm{x},t)+g\delta^{d}(\bm{x}).
\end{equation}
Unlike the mean-field equation for the model with competitions, this equation can be solved exactly in all dimensions.  If we assume an initial condition with no particles present, a Laplace transform in time and Fourier transform in space leads to:
\begin{equation}
\widetilde{\bar{n}}(\bm{k},s)=\frac{g}{s(s+D\bm{k}^{2}+z)}.
\end{equation}
Transforming back into the time domain gives:
\begin{equation}
\widetilde{\bar{n}}(\bm{k},t)=\frac{g\left[1-e^{-(z+D\bm{k}^{2})t}\right]}{D\bm{k}^{2}+z}.
\end{equation}
We are interested in the long-time, steady-state behavior in all dimensions.  Letting $t\to\infty$ and transforming in space gives the following solutions for $\bar{n}_{ss}(\bm{x})\equiv\bar{n}(\bm{x},t=\infty)$:
\begin{eqnarray}
\bar{n}_{ss}(x) & = & \frac{ge^{-\kappa |x|}}{\sqrt{4Dz}}\qquad1\text{D}\nonumber\\
\bar{n}_{ss}(r) & = & \frac{gK_{0}(\kappa r)}{4\pi D}\qquad2\text{D}\nonumber\\
\bar{n}_{ss}(r) & = & \frac{ge^{-\kappa r}}{4\pi Dr}\qquad3\text{D}\label{MFlinsols}
\end{eqnarray}

There is one additional case of interest: the $d=1$ lattice case.  The relevant mean-field equation in this case is simply (\ref{MFlinwsource}).  After a Laplace transform, we are left with a difference equation which can be solved with the ansatz $\widetilde{\bar{n}}(\nu+1,s)=e^{-f(s)}\widetilde{\bar{n}}(\nu,s)$ for $\nu>0$.  The solution is
\begin{equation}
\widetilde{\bar{n}}(\nu,s)=\frac{ge^{-f(s)|\nu|}}{sw\sinh(f(s))},
\end{equation}
where $f(s)=\cosh^{-1}(1+(s+z)/w)$.  We can immediately get the $t\to\infty$ behavior of $\bar{n}(\nu,t)$ from this expression by multiplying by $s$ and letting $s\to0$, resulting in
\begin{equation}
\bar{n}_{ss}(\nu)\equiv\bar{n}(\nu,t\to\infty)=\frac{ge^{-f|\nu|}}{w\sinh(f)},
\end{equation}
where $f\equiv f(0)$.

The functional forms of the continuum solutions in (\ref{MFlinsols}) are the same as those of the  solutions for the asymptotic ($r\gg a$) steady-state nonlinear ($b\neq0$) equations discussed in Section \ref{MF1Oasis}.  For a properly chosen creation rate $g$, the mean-field solutions of the two models should match at long distances.  We will use this method of matching mean-field solutions to determine $g$ for the purposes of making numerical predictions of first passage properties in the nonlinear model.  It is important to note that $g$ is not a ``fit parameter'': its value is completely determined by the oasis size, the death rate, etc., and is not adjusted to fit data generated by the nonlinear model.

In practice, one can solve the nonlinear steady-state mean-field equations numerically, and then find $g$ by matching the long-distance behavior to the appropriate solution from (\ref{MFlinsols}).  It is possible, however, to match the $d=1$ solutions analytically: using the results of Section \ref{MF1Oasis} together with (\ref{MFlinsols}) results in
\begin{equation}
g=8\sqrt{Dz}\,\gamma^{2}\bar{c}_{ss}(a)e^{\kappa a},
\end{equation}
where $\gamma^{-1}=1+\sqrt{2q\bar{c}_{ss}(a)/3z}$, as before.  The constant $\bar{c}_{ss}(a)$ can be found as described in Section \ref{MF1Oasis}.  As $y\to\infty$, $g\to12\sqrt{Dz^3}\,e^{\kappa a}/q$.  For higher dimensions, it is necessary to numerically solve the mean-field equations for the nonlinear model to accurately calculate $g$.


\subsection{\label{preds}Analytic Predictions from the Linear Model with a Source}
With a method in place for determining $g$ from the parameters of the nonlinear model, it is now possible to use the linear model with a source to make predictions about first passage properties of the two oasis system.  We begin by noting that, since the particles in the linear model with a source are non-interacting, the full multi-particle FPT PDF $f_{N}(\bm{x},t)$---that is, the probability per unit time that the first particle from the first oasis reaches the second oasis between $t$ and $t+dt$---can be written in terms of the one-particle FPT PDF $f_{1}(\bm{x},t)$.  (Note that the vector $\bm{x}$ is a stand-in for all the geometric particulars of the system.  For instance, for a spherical or circular oasis, $f_{N}(\bm{x},t)$ depends on the distance of the center of the oasis from the origin $R$ and the radius $a$ of the oasis.  These geometrical particulars are not important for our present discussion, and so we express $f_{N}$ as a function of the generic vector $\bm{x}$.)  This is accomplished as follows:  assume the source is at the origin, and that it releases $N$ particles per unit time $\Delta t$ \cite{noteonPoisson}.  Define $S(\bm{x},t)=1-\int_{0}^{t}dt'\,f_{1}(\bm{x},t')=1-P_{\text{hit}}(\bm{x},t)$ to be the probability that a particular particle released from the origin at $t=0$ has \emph{not} reached the target oasis by time $t$.  If we define $P_{\text{none}}(\bm{x},t)$ to be the probability that \emph{no} particles from the source have hit the target oasis by time $t$, then
\begin{equation}
P_{\text{none}}(\bm{x}, t)\,=\prod_{\tau=0, \Delta t, \ldots}^{t}[S(\bm{x},\tau)]^{N}.\label{probnone1}
\end{equation}
Taking the logarithm of this expression gives
\begin{equation}
\ln\left[P_{\text{none}}(\bm{x},t)\right]=\sum_{\tau=0, \Delta t, \ldots}^{t}g\Delta t\,\ln\left[S(\bm{x}, \tau)\right],\label{probnonelog}
\end{equation}
where $g\equiv N/\Delta t$ is the creation rate.  Taking the limit $\Delta t\to0$ with $g$ fixed and exponentiating both sides leads to a closed equation for $P_{\text{none}}(\bm{x},t)$ in terms of $S(\bm{x},t)$:
\begin{equation}
P_{\text{none}}(\bm{x}, t)=\exp\left(g\int_{0}^{t}dt'\,\ln S(\bm{x}, t')\right).
\end{equation}
Since we are interested in oasis separations large enough that a given single particle has a low probability of ever reaching the second oasis, $S(\bm{x},t)$ is close to $1$ even as $t\to\infty$.  This allows us to approximate $\ln S(\bm{x},t)=\ln (1-P_{\text{hit}}(\bm{x},t))$ by $-P_{\text{hit}}(\bm{x},t)$, leading to a simpler expression for $P_{\text{none}}(\bm{x},t)$:
\begin{equation}
P_{\text{none}}(\bm{x}, t)\simeq\exp\Big[-g\int_{0}^{t}dt'\, (t-t')f_{1}(\bm{x}, t')\Big]. \label{finalpnone}
\end{equation}
The full FPT PDF $f_{N}(\bm{x},t)$ is simply $-\partial_{t}P_{\text{none}}(\bm{x},t)$.  

There is one more useful way to write $P_{\text{none}}$: since the integral appearing in the exponent in (\ref{finalpnone}) is a convolution of $t$ and $f_{1}(\bm{x},t)$, its Laplace transform is simply a product of the two functions' individual Laplace transforms.  Explicitly:
\begin{equation}
P_{\text{none}}(\bm{x},t)\simeq\exp\left(-g\mathcal{L}^{-1}\left[\widetilde{f}_{1}(\bm{x},s)/s^2\right]\right),\label{pnonelap}
\end{equation}
where $\mathcal{L}^{-1}[u(s)]$ is the inverse Laplace transform of $u(s)$ and $\widetilde{f}_{1}(\bm{x},s)$ is the Laplace transform in time of $f_{1}(\bm{x},t)$.  Often it is easier to compute $\widetilde{f}_{1}(\bm{x},s)$ than $f_{1}(\bm{x},t)$, and in these cases (\ref{pnonelap}) can be very useful.

In order to make predictions using (\ref{finalpnone}) or (\ref{pnonelap}), it is necessary to compute the one-particle FPT PDF $f_{1}(\bm{x},t)$.  We will do this now for the continuum case in all relevant dimensions and the lattice case in $d=1$.  We will start with the continuum case.  The diffusion equation governing the probability distribution $p_{1}(\bm{x},t)$ of a particle released into the desert from the origin at $t=0$ is
\begin{equation}
\frac{\partial p_{1}(\bm{x},t)}{\partial t}=D\nabla^{2}p_{1}(\bm{x},t)-zp_{1}(\bm{x},t),\label{1pdiff}
\end{equation}
with boundary condition $p_{1}(\text{oasis surface},t)=0$.  This boundary condition is of course not true in the model---particles arriving at the oasis will not immediately die---but it is used as a device to extract first passage properties.  By writing $p_{1}(\bm{x},t)=\phi_{1}(\bm{x},t)e^{-zt}$, it is possible to eliminate the death term in (\ref{1pdiff}) and arrive at a simple diffusion equation for $\phi_{1}(\bm{x},t)$.  The FPT PDF $f_{1}(\bm{x},t)$ can be obtained by considering the flux of probability into the oasis \cite{rednerbook}:
\begin{equation}
f_{1}(\bm{x},t)=D\int_{\substack{\text{oasis} \\ \text{surface}}} dA\,\,\hat{n}\cdot\nabla\phi_{1}(\bm{x},t)e^{-zt},\label{1partflux}
\end{equation}
where $dA$ is an element of the oasis surface and $\hat{n}$ is a unit vector pointing out from the oasis.  Since $\phi_{1}(\bm{x},t)$ is the solution to a simple diffusion equation, $D\int dA\,\hat{n}\cdot\nabla\phi_{1}(\bm{x},t)=f_{1}^{z=0}(\bm{x},t)$, the FPT PDF in the case where there is no desert.  This fact can be combined with (\ref{1partflux}) to arrive at the conclusion
\begin{equation}
f_{1}(\bm{x},t)=f_{1}^{z=0}(\bm{x},t)e^{-zt}.\label{f1z0relation}
\end{equation}
The Laplace-transformed FPT PDF $\widetilde{f}_{1}(\bm{x},s)$ is thus related to the $z=0$ function by
\begin{equation}
\widetilde{f}_{1}(\bm{x},s)=\widetilde{f}_{1}^{z=0}(\bm{x},s+z).\label{z2nozfreq}
\end{equation}
These results are convenient due to the fact that, for circular or spherical oases, exact solutions exist for $f_{1}^{z=0}(\bm{x},t)$.

\begin{figure}
\includegraphics[scale=0.35]{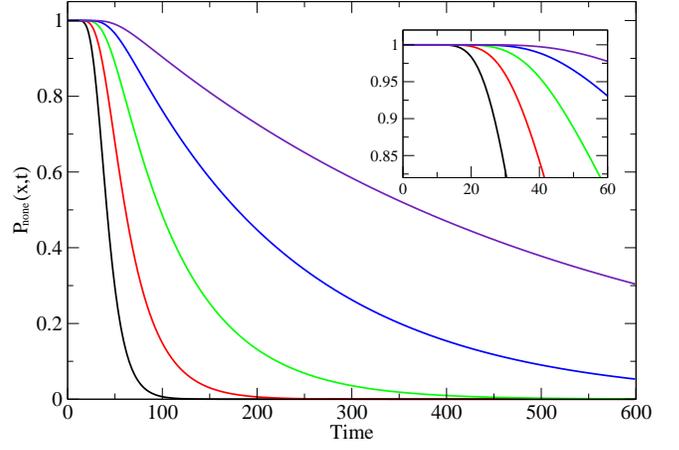}
\caption{\label{Pnone1Dfig}Main window: plot showing $P_{\text{none}}(x,t)$ in $d=1$.  The lines represent, from left to right, the function for $x=16,18,20,22,$ and $24$.  Inset: a blowup showing the early-time behavior of $P_{\text{none}}$.}
\end{figure}

In one dimension, $f_{1}^{z=0}(x,t)=|x|e^{-x^{2}/4Dt}/\sqrt{4\pi Dt^{3}}$ \cite{rednerbook}.  This means that
\begin{equation}
f_{1}(x,t)=\frac{|x|e^{-x^{2}/4Dt}e^{-zt}}{\sqrt{4\pi D}\,t^{3/2}}\label{f11Dcont}
\end{equation}
when there is a desert present.  Plugging this into (\ref{finalpnone}) and doing the integration \cite{abramandstegun} gives
\begin{eqnarray}
P_{\text{none}}(x,t) & \simeq & \exp\left[-\frac{g}{4z}\left(e^{\kappa |x|}\,\zeta^{+}\,\text{erfc}(\zeta^{+}/\sqrt{4zt}\,)\right.\right.\nonumber\\
& & \left.\left.-e^{-\kappa |x|}\,\zeta^{-}\,\text{erfc}(\zeta^{-}/\sqrt{4zt}\,)\right)\right],\label{Pnone1Dcont}
\end{eqnarray}
where $\zeta^{\pm}=\zeta^{\pm}(x,t)=\kappa |x|\pm 2zt$.  This function is shown in Fig. \ref{Pnone1Dfig}.  For large times, $P_{\text{none}}(x,t)\sim\exp\left(-ge^{-\kappa |x|}t\right)$.  The $j$-th moment of $f_{N}(x,t)$ is given by $\langle T^{j}(x)\rangle=j\int_{0}^{\infty}dt\,P_{\text{none}}(x,t)t^{j-1}$; although it is not possible to perform this integral analytically, we can extract its $|x|\to\infty$ (large oasis separation) behavior (see Appendix \ref{asymp}):
\begin{equation}
\langle T^{j}(x)\rangle=j!\,\frac{e^{\kappa |x|j}}{g^{j}}.\quad{(1\text{D continuum})}\label{1Dcontmoms}
\end{equation}

In two and three dimensions, it becomes more convenient to solve for $\widetilde{f}_{1}(\bm{x},s)$ and then use (\ref{pnonelap}) to obtain $P_{\text{none}}$.  The single-particle FPT PDF is a function of the separation of the center of the target oasis from the origin ($\bm{R}$) and the radius of the oasis ($a$), so we will from now on write it as $f_{1}(R,a,t)$, where $R=|\bm{R}|$.  The FPT PDF in frequency space in the absence of a desert ($z=0$) is known for these cases \cite{rednerbook}; using (\ref{z2nozfreq}) gives
\begin{equation}
\widetilde{f}_{1}(R,a,s)=\left(\frac{a}{R}\right)^{d/2-1}\,\frac{K_{d/2-1}\left(\sqrt{\frac{s+z}{D}}\,R\right)}{K_{d/2-1}\left(\sqrt{\frac{s+z}{D}}\,a\right)},\label{f1lap}
\end{equation}
where $K_{n}$ is the $n$-th order modified Bessel function of the first kind.  This equation also holds in $d=1$; redefining $x=R-a$ and using the definition of $K_{1/2}$ leads to the Laplace transform of (\ref{f11Dcont}).

In $d=2$, using (\ref{pnonelap}) and (\ref{f1lap}) gives
\begin{equation}
P_{\text{none}}(R,a,t)\simeq\exp\left[-\frac{g}{2\pi\imath}\int_{\mathcal{L}}ds\,\frac{e^{st}K_{0}\left(\sqrt{\frac{s+z}{D}}\,R\right)}{s^{2}K_{0}\left(\sqrt{\frac{s+z}{D}}\,a\right)}\right].
\end{equation}
The exponent can be reduced to an analytic function plus a real integral (see Appendix \ref{asymp})
\begin{eqnarray}
P_{\text{none}}(R,a,t) & = & e^{-gY(R,a,t)}\nonumber\\
Y(R,a,t) & =  & \frac{t\,K_{0}\left(\kappa R\right)}{K_{0}\left(\kappa a\right)}\label{2DFPTPDF}\\
& - & \frac{RK_{0}\left(\kappa a\right)K_{1}\left(\kappa R\right)-aK_{0}\left(\kappa R\right)K_{1}\left(\kappa a\right)}{\sqrt{4Dz}\left[K_{0}\left(\kappa a\right)\right]^{2}}\nonumber\\
& + & \frac{2R^{2}e^{-zt}}{\pi D}\int_{0}^{\infty}du\,\frac{u\,e^{-Dtu^{2}/R^{2}}\,Z_{0}(u,\frac{a}{R})}{\left[u^{2}+\kappa^{2}R^{2}\right]^{2}},\nonumber
\end{eqnarray}
where 
\begin{equation}
Z_{0}\left(u,\frac{a}{R}\right)=\frac{J_{0}(\frac{a}{R}u)Y_{0}(u)-J_{0}(u)Y_{0}(\frac{a}{R}u)}{J_{0}(\frac{a}{R}u)^{2}+Y_{0}(\frac{a}{R}u)^{2}}.\label{Zdef}
\end{equation}
The large $t$ behavior of $P_{\text{none}}$ is given by $P_{\text{none}}(R,a,t)\sim\exp\left(-g\left[K_{0}(\kappa R)/K_{0}(\kappa a)\right]t\right)$.  The moments asymptotically approach
\begin{equation}
\langle T^{j}(R,a)\rangle=j!\left(\frac{K_{0}(\kappa a)}{gK_{0}(\kappa R)}\right)^j\quad{(2\text{D contin.})}\label{2Dcontmoms}
\end{equation}
as $R\to\infty$.

The three dimensional case is easy to treat.  Since $K_{-n}(z)=K_{n}(z)$, looking at (\ref{f1lap}) immediately shows that $\widetilde{f}_{1}(R,a,s)$ for $d=3$ is identical to the $d=1$ case save for a factor of $a/R$.  Making the replacements $|x|\to R-a$ and $g\to ga/R$ in (\ref{Pnone1Dcont}) gives $P_{\text{none}}(R,a,t)$; making the same replacements gives the $t\to\infty$ decay $P_{\text{none}}(R,a,t)\sim\exp\left(-g(a/R)e^{-\kappa(R-a)}t\right)$.  The moments approach
\begin{equation}
\langle T^{j}(R,a)\rangle=j!\left(\frac{R}{a}\right)^{j}\frac{e^{\kappa(R-a)j}}{g^{j}}\quad{(3\text{D contin.})}\label{3Dcontmoms}
\end{equation}
as $R\to\infty$.

The final case we will consider is the $d=1$ lattice case.  Recall that for this case, $w$ is total hopping rate and the integer $\nu$ denotes the lattice point.  The single-particle FPT PDF $f_{1}(\nu,t)$ is \cite{rednerbook}
\begin{equation}
f_{1}(\nu,t)=\frac{|\nu|e^{-(w+z)t}I_{\nu}(wt)}{t},
\end{equation}
where $I_{\nu}$ is the $\nu$-th order modified Bessel function of the first kind.  It is more convenient to use the frequency space function:
\begin{equation}
\widetilde{f}_{1}(\nu,s)=\frac{w^{|\nu|}}{\left[s+w+z+\sqrt{(s+z)(s+z+2w)}\right]^{|\nu|}}
\end{equation}
Using this together with (\ref{pnonelap}) gives an expression for $P_{\text{none}}(\nu,t)$ (see Appendix \ref{asymp}):
\begin{eqnarray}
P_{\text{none}}(\nu,t) & \simeq & e^{-gY(\nu,t)}\nonumber\\
Y(\nu,t) & = & e^{-f|\nu|}\left(t-\frac{|\nu|}{w\sinh(f)}\right)\\
& + & \frac{e^{-(w+z)t}}{\pi w}\int_{0}^{\pi}d\theta\,\frac{\sin(\theta)\sin(|\nu|\theta)e^{wt\cos(\theta)}}{\left[1+\frac{z}{w}-\cos(\theta)\right]^2}.\nonumber
\end{eqnarray}
This function decays as $P_{\text{none}}(\nu,t)\simeq\exp\left(-ge^{-f|\nu|}t\right)$ as $t\to\infty$.  As in the continuum case, $P_{\text{none}}$ cannot be integrated analytically, but an asymptotic analysis (see Appendix \ref{asymp}) shows that, as $|\nu|\to\infty$,
\begin{equation}
\langle T^{j}(\nu)\rangle=j!\,\frac{e^{f|\nu|j}}{g^{j}}.\quad{(1\text{D lattice})}\label{1Dlattmoms}
\end{equation}

\subsection{Simulation Results}
In order to test the predictions of the linear theory with a source, we wrote a kinetic Monte Carlo simulation of the model with interactions.  While it is certainly possible to simulate the continuum model in any dimension either by doing a discrete-space simulation and choosing very small lattice spacings or by using an event-driven algorithm \cite{donevproc}, we found it more expedient to do a lattice simulation in $d=1$ and compare with the predictions from the lattice version of the linear model with a source.

Each simulation run began with a population of $y/2b=125$ particles sitting in the middle site of a $5$-lattice-point-wide oasis.  Once a given end point was reached for the first time, the run ended.  In order to minimize sampling error, $5000$ runs were performed.

\begin{figure}
\includegraphics[scale=0.35]{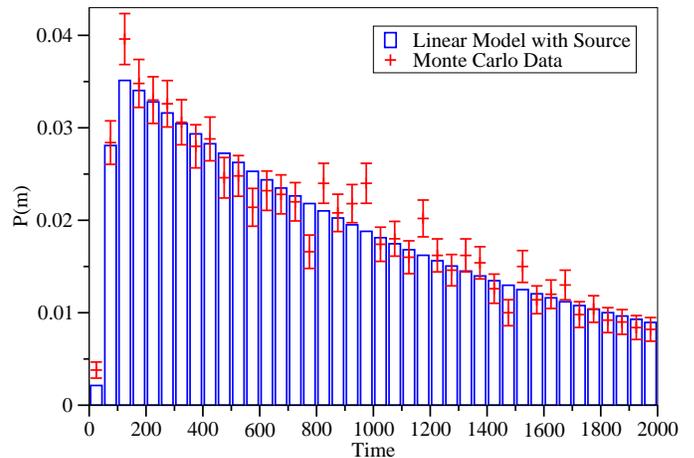}
\caption{\label{histo}Binned FPT probabilities for $\nu=27$ from both the linear model with a source (blue boxes) and Monte Carlo simulations of the model with interactions (red lines).  The width of each bin is $50/w$, where $w$ is the total hopping rate.  The error bars on the simulation data represent sampling error.  Only times up to $t=2000$ are shown for the sake of clarity.}
\end{figure}

The agreement between the predictions from the linear model with a source and the Monte Carlo simulation results from the model with interactions is excellent.  The linear model with a source correctly predicts the lower moments of $f_{N}(\nu,t)$ for large $\nu$, as shown in Table\ \ref{momenttable}.  A more stringent test of the power of the linear model with a source is a comparison of its prediction for the full FPT PDF with simulation results.  To do this comparison, we integrated $f_{N}(\nu,t)$ from $(m-1)\Delta t$ to $m\Delta t$ for $m=1,2,3\ldots M$ to obtain a set of probabilities $P(\nu,m)$ for hitting the point $\nu$ for the first time in time bin $m$.  We then compared this prediction with simulation results.  The comparison is shown in Fig.\ \ref{histo} for $\nu=27$; it seems clear that the linear model with a source correctly predicts the form for $f_{N}(\nu,t)$.

\begin{table*}
\caption{\label{momenttable}Comparison of predictions from the linear model with a source for the first, second, and third moments with Monte Carlo data from the model with interactions.  The quoted errors represent a 95\% confidence interval.}
\begin{ruledtabular}
\begin{tabular}{|c|c|c|c|c|c|c|}
Distance & $\langle T^{1}\rangle_{\text{th}}$ & $\langle T^{1}\rangle_{\text{sim}}$ & $\langle T^{2}\rangle_{\text{th}}$ & $\langle T^{2}\rangle_{\text{sim}}$ & $\langle T^{3}\rangle_{\text{th}}$ & $\langle T^{3}\rangle_{\text{sim}}$\\\hline
$\nu=10$ & $12.7781$ & $12.4059\pm.092208$ & $172.986$ & $164.97\pm2.38404$ & $2464.79$ & $2328.16\pm50.5446$\\
$\nu=15$ & $33.0596$ & $33.3945\pm.288245$ & $1196.65$ & $1223.91\pm21.8565$ & $47264.1$ & $48850.9\pm1434.07$\\
$\nu=20$ & $102.398$ & $103.966\pm1.72726$ & $14321.6$ & $14691.1\pm572.021$ & $2.69537\times10^{6}$ & $(2.74028\pm.195672)\times10^{6}$\\
$\nu=25$ & $609.336$ & $612.667\pm15.7994$ & $6.79632\times10^{5}$ & $(7.00186\pm.424694)\times10^{5}$ & $1.13187\times10^{9}$ & $(1.21234\pm.148632)\times10^{9}$\\
$\nu=30$ & $5164.80$ & $5066.48\pm140.589$ & $5.26790\times10^{7}$ & $(5.13892\pm.321111)\times10^{7}$ & $8.05889\times10^{11}$ & $(7.88352\pm.918500)\times10^{11}$\\
\end{tabular}
\end{ruledtabular}
\end{table*}

\section{\label{ManyOases}From Two Oases to Many}
\subsection{\label{hopping}The Connection with Hopping Conduction}We have shown that the first passage time statistics of the two-oasis model with competition ($2A\to A$) are adequately captured by a simple solvable model without competition when the oasis separation $R$ is large.  We would like to apply these results to systems with more than two oases in order to determine the nature of of transport in a large system.

For concreteness, consider a continuum system in $d$ dimensions ($d>1$) comprised of identical oases of radius $a$ and growth rate $y$ placed around randomly distributed points with number density $n$ in a desert of death rate $z$.  We are interested in the low density regime; that is, the regime in which the average distance between oases is larger than the lengthscale $R_{\text{lin}}$ identified in (\ref{Rlin1D}) \cite{noteonlengthscale}.  We will allow the oases to overlap, although this shouldn't happen too often at the low oasis densities we are considering.  We will start with one or more oases populated at $t=0$ and wait for a particular oasis or one of a number of oases situated far away to become populated.  We will call the total time for this to take place  $T_{\text{infection}}$, the infection time.  Because of the exponential dependence of the mean FPT on oasis separation for large oasis separations (see (\ref{1Dcontmoms}), (\ref{2Dcontmoms}), and (\ref{3Dcontmoms})), the time taken to cross the largest oasis separations (or links) on the path should, on average, be \emph{much} greater than the time taken to cross the shorter links.  The situation is somewhat analogous to that of hopping conduction in doped semiconductors \cite{shklovskii}: the oases in our system play the role of the impurity sites in the semiconductor, and the mean transit time between oases is akin to the resistance between impurity sites.  In doped semiconductors, the resistance between impurity sites depends exponentially on their separation like $e^{\alpha R}$, where $R$ is the impurity separation and $\alpha\equiv2/a$, where $a$ is an effective Bohr radius describing the width of the impurity wavefunctions \cite{shklovskii}.  This is similar to the way the mean transit time (and, indeed, all other moments of the distribution for large separation) depends exponentially on oasis separation in our system.  There are a couple of significant differences between the two systems: first, there is no equivalent in the semiconductor problem of the growth time, the time needed for the population on a newly inhabited oasis to rise to a significant level; second, the resistances between impurity sites are not the averages of stochastic variables like the mean transit times, but rather definite quantities.  The first of these differences is insignificant since we have already assumed that $T_{\text{growth}}$ is much smaller than a typical value of  $T_{\text{transit}}$ for oases separated by a large distance.  The second difference is more important, and some of its implications will be discussed in detail later in this paper.

The problem of determining the resistivity (or conductivity) of a doped semiconductor in the hopping regime was first tackled satisfactorily using ideas from percolation theory by Ambegaokar and coworkers \cite{ambegaokarperc,kurkijarvi74}.  They found that the resistivity is dominated by the largest links in the network of impurity sites spanning the system.  Any links with much larger resistances are effectively shunted by the smaller resistances, and are not important in determining the macroscopic resistivity.  The size of the largest link $R_{\text{max}}$ can be determined using continuum percolation theory, which works in roughly the following way: a circle (or sphere) is drawn around each impurity site, and the radius of each circle is increased.  When an impurity site center comes within the circle centered around another impurity site, the two are said to be linked.  When the radii of the circles are increased to the point where a cluster of linked sites connects one side of the system to another, we have reached the percolation threshold.  The last link to form is clearly the longest link, and we call its length $R_{\text{max}}$.  This length varies from sample to sample, but has a well-defined limit as the system size goes to infinity \cite{shklovskii}:
\begin{equation}
R_{\text{max}}=\left[\frac{B_{c}(d)}{n\,V_{d}}\right]^{1/d},\label{Rmax}
\end{equation}
where $B_{c}(d)$ is the dimensionally-dependent \emph{bonding criterion}, $V_{d}$ is the volume of a $d$-dimensional unit hypersphere, and $n$ is the number density of impurity sites.  The quantity $B_{c}(d)$ has an interpretation as the mean number of connected neighbors for members of the percolation cluster \cite{shklovskii}.

The network which carries the majority of the current in doped semiconductors is called the critical subnetwork, and its correlation length is $L_{0}$ (this is also the length scale at which sample-to-sample variations in $\alpha R_{\text{max}}$ become relatively small, of order $1$) \cite{shklovskii}.  Above this length scale, the system can be regarded as homogeneous, and so the resistivity of a large system of size $L\gg L_{0}$ is roughly equal to the resistivity of a system of size $L_{0}$.  As argued above, this resistivity is largely determined by the resistance across the largest link, which is equal to $e^{\alpha R_{\text{max}}}/G_{0}$ ($G_{0}$ has units of conductance and $\alpha$ is equal to the inverse).  The resistivity $\rho$ is then given by \cite{kurkijarvi74,shklovskii}
\begin{equation}
\rho\simeq\frac{e^{\alpha R_{\text{max}}}{L_{0}}^{d-2}}{G_{0}},\label{resistivity}
\end{equation}
with the correlation length $L_{0}$ given approximately by
\begin{equation}
L_{0}\simeq\frac{\left(\alpha R_{\text{max}}\right)^{\nu}}{n^{1/d}},\label{corr-semi}
\end{equation}
where $n$ is the number density of impurity sites and $\nu$ is a critical exponent equal to $4/3$ in $d=2$ and $\sim .9$ in $d=3$.

\subsection{\label{onecorrlen}Dynamics of Transport in a Macroscopic System}
Now let us return to our problem.  Consider a system of the size of the correlation length $L_{0}$ of the subcritical network with one oasis initially infected at one edge of the system.  In the hopping conduction problem, the goal is to find the resistance between the edges of the system; in our problem, it is to find the first passage time between the starting oasis and either a specific oasis on the opposite edge or any oasis in a thin layer close to the opposite edge.  Unlike the hopping conduction problem, our problem is dynamic in nature; an additional difference is that, as mentioned previously, our first passage times are random variables with a distribution whose mean increases exponentially with oasis separation rather than fixed resistances with an exponential dependence on link size.  The mean FPT across the system is thus an average of a minimum: for a fixed set of oases, each realization of the dynamic process yields a path with minimal first passage time which may differ from the paths from other realizations.  However, there is at least one large link of size $\simeq R_{\text{max}}$ which must be crossed in order for the population to reach the opposite edge of the system, and the time to cross this link sets the time scale to cross the system in the same way the that the resistance of the largest link sets the scale of the resistance in the hopping conduction problem.  Thus,
\begin{equation}
\langle\text{time to cross system of size }L_{0}\rangle\simeq\langle T(R_{\text{max}},a)\rangle,\label{onecorreqn}
\end{equation}
where $T(R_{\text{max}},a)$ is given by (\ref{2Dcontmoms}) or (\ref{3Dcontmoms}) depending on the dimensionality of the system.

Now consider a very large system.  We wish to find the mean infection time $\langle T_{\text{infection}}(L)\rangle$---that is, the mean time for the population to travel between oases separated by some large distance $L\gg L_{0}$.  This time is roughly equal to the mean FPT in the parameter regime in which we're interested (that is, the limit of high growth rate on the oases).  In order to do this, we need to know something about the large-scale structure of the cluster of oases which will carry the bulk of the particle current.  Again, looking at the hopping conduction problem is instructive.  In that problem, the links-nodes-blobs picture \cite{shklovskii,percolbook} suggests that the current-carrying cluster can be thought of as a network of nodes separated by a distance on the order of $L_{0}$ connected by one-dimensional links and clusters (or blobs) of links.  Since the resistance of a link depends exponentially on its length, the largest one-dimensional links of approximate size $R_{\text{max}}$ largely determine the resistance between nodes, and thus the resistivity of the system, as noted in the previous section.  (There is some debate as to whether there exists another length scale $l$ which, together with $L_{0}$, characterizes the structure of the current-carrying cluster.  See Ref. \cite{hunt2005ptf} for a discussion of this problem.)

As a first approximation, let us consider our system as consisting of nodes placed on a hypercubic lattice with lattice spacing $L_{0}$ with one large link of size $R_{\text{max}}$ in between each node.  We ignore the time to cross the shorter links and the variations in the oasis configurations from one correlation-length-sized chunk to another.  The population starts at one node, and we seek the first passage time to some distant node located a distance $L$ away along a lattice basis vector (or, equivalently, $n=L/L_{0}$ lattice points away).  This is the basic problem of \emph{first passage percolation} (FPP), a field largely studied in the mathematical community \cite{kesten1986afp}.  One of the basic results of FPP is that, as the separation between nodes $n\to\infty$, the FPT $\langle T(n)\rangle$ divided by $n$ goes to a constant $\mu$, conventionally called the time constant.  Thus, the mean FPT rises linearly with distance between sites, indicating that the proper intensive quantity for our problem is the mean FPT divided by oasis separation; in the doped semiconductor problem, the proper intensive quantity is the resistivity.  The value of $\mu$ depends on the underlying FPT probability distribution, but a general result is that $\mu\leq\langle T_{1}\rangle$, where $\langle T_{1}\rangle$ is the average time to cross one link \cite{kesten1986afp}.  (That is, $\langle T_{1}\rangle$ is the mean of the distribution from which FPTs are picked for each link between nodes.)  For the case where the times are chosen from an exponential distribution, $\mu\simeq.4\langle T_{1}\rangle$ in two dimensions \cite{alm2002lau}.  In general, $\langle T_{1}\rangle$ is an upper limit on $\mu$ \cite{kesten1986afp}.

Since we are interested in obtaining a rough estimate of the infection time, we will simply use the upper limit $\langle T_{1}\rangle$ (the mean time to cross one link) as an estimate for $\langle T_{\text{infection}}\rangle/n$.  This gives us the following:
\begin{equation}
\frac{\langle T_{\text{infection}}(L)\rangle}{L}\simeq\frac{\langle T(R_{\text{max}},a)\rangle}{L_{0}}.\label{tinfresult}
\end{equation}
where $\langle T(R_{\text{max}},a)\rangle$ is again given by (\ref{2Dcontmoms}) in $d=2$ and (\ref{3Dcontmoms}) in $d=3$, and $L_{0}$ is given by (see (\ref{corr-semi}))
\begin{equation}
L_{0}\simeq\frac{\left(\kappa R_{\text{max}}\right)^{\nu}}{n^{1/d}}.\label{corr-fpp}
\end{equation}
This result is an order-of-magnitude estimate, but it should capture the dependence of $\langle T_{\text{infection}}\rangle$ on the relevant parameters of the system.  

It is probably good to stop at this point and briefly recall the approximations that we have made to obtain the result in (\ref{tinfresult}): first, we have ignored the growth time on the grounds that it is small compared to the transit time between oases; second, we have simplified the picture of transport on the scale of $L_{0}$, replacing the mess of oases with a single link of size $R_{\text{max}}$; and third, we have used an upper limit on the time constant rather than the time constant itself.  It should be noted that the first and second approximations tend to lead to underestimating $\langle T_{\text{infection}}\rangle$, while the third tends to lead to overestimating it.

\subsection{\label{largesims}Comparison with Simulations}
In order to confirm the predictions of the preceding section, we wrote a program capable of simulating a very large system in two dimensions.  To make the simulation of such a large system tractable, we made some important simplifications which must be explained.  The first of these is the most important: rather than simulating the motion of individual particles, we simply assigned first passage times between oases.  This allowed us to go to system sizes many orders of magnitude larger than we could have achieved via a full kinetic Monte Carlo simulation involving every particle.  

The second simplification involves the nature of the FPT PDF used to generate the passage times between oases.  The linear theory with a source produces an analytical expression for this FPT PDF (see Eqs. (\ref{2DFPTPDF}) and (\ref{Zdef})), but this is unwieldy and computationally expensive to calculate.  However, for large $R$, the moments of this FPT PDF in $d=2$ approach those of an exponential distribution with parameter $gK_{0}(\kappa R)/K_{0}(\kappa a)$ (see \ref{2Dcontmoms}), where $\kappa\equiv\sqrt{z/D}$.  Since it is the large-$R$ separations which will largely determine the infection time, we simply replaced the complicated FPT PDF between oases with this exponential distribution; the errors introduced by this simplification are serious only for small oasis separations, and these do not contribute much to the infection time.

The remaining simplifications are minor: we treated all the oases as points; we ignored the growth time, just as we have done in the analytical work presented in the preceding sections; and finally, we ignored the effects of neighboring oases on the first-passage time statistics between two oases.  This final simplification again introduces errors mostly in areas of high oasis density where oasis separations are small.  The bottlenecks of our particle current-carrying cluster occur where there are two oases separated by a large region of desert, and in these areas the FPT statistics should be very close to those derived in the case of two oases in an infinite desert.

Before presenting our simulation results, we must first provide some details of the way time was scaled in our simulations.  With the simplifications we have made, the FPT PDF $f_{N}(R,a,t)$ between a pair of oases of radii $a$ separated by a distance $R$ can be obtained from (\ref{2DFPTPDF}).  It is given by:
\begin{equation}
f_{N}(R,a,t)=\frac{gK_{0}\left(\kappa R\right)}{K_{0}\left(\kappa a\right)}\,\exp\left[-\frac{gK_{0}\left(\kappa R\right)}{K_{0}\left(\kappa a\right)}\,t\right].
\end{equation}
If we define the variable $\tau=gtK_{0}(\kappa R_{\text{max}})/K_{0}(\kappa a)$---effectively measuring time in units of the mean time to cross a link of size $R_{\text{max}}$---we can absorb the dependence of the FPT PDF on $a$ and $g$ into $\tau$.  The FPT PDF then becomes
\begin{equation}
f_{N}(R,\tau)=\frac{K_{0}(\kappa R)}{K_{0}(\kappa R_{\text{max}})}\,\exp\left[-\frac{K_{0}(\kappa R)}{K_{0}(\kappa R_{\text{max}})}\,\tau\right].\label{FPTfortau}
\end{equation}
Our simulation measured time in units of $\tau$, so that there was no need to input information about $g$ or the oasis size $a$.

There is one further approximation that we made in our simulations simply for the sake of convenience: we used the large-argument asymptotic form for $K_{0}(x)$ of $\sqrt{\pi/2x}\,e^{-x}$, making the FPT PDF
\begin{equation}
f_{N}(R,\tau)=\sqrt{\frac{R_{\text{max}}}{R}}\,e^{-\kappa(R-R_{\text{max}})}\,\exp\left[-\sqrt{\frac{R_{\text{max}}}{R}}\,e^{-\kappa(R-R_{\text{max}})}\,\tau\right].\label{FPTfortausimp}
\end{equation}
Like some of the other simplifications and approximations we made in the simulations, this approximation is not good for small oasis separations, but the errors introduced are ultimately unimportant given the contribution of the small oasis jumps to the transit time.

If our theory is correct, the mean time to cross one block of size $L_{0}$ in these units (in units of $\tau$) should be of order $1$, and the mean infection time should be
\begin{equation}
\langle\tau_{\text{infection}}(L)\rangle\simeq\left(\frac{L}{L_{0}}\right).\label{tauinf1}
\end{equation}
If $\kappa$ and $R_{\text{max}}$ are adjusted in such a way so that their product remains constant, then this amounts to a trivial rescaling of space, and $\tau_{\text{infection}}$ should simply vary as $1/R_{\text{max}}$.  This is already captured through the dependence of $\tau_{\text{infection}}(L)$ on $L_{0}$, and so we can rewrite (\ref{tauinf1}) as
\begin{equation}
\langle\tau_{\text{infection}}(L)\rangle=\left(\frac{L}{L_{0}}\right)F(\kappa R_{\text{max}}),\label{tauinf2}
\end{equation}
where $F(\kappa R_{\text{max}})$ is some function of order unity.  We thus expect that a graph of $\langle\tau_{\text{infection}}\rangle$ versus $L/L_{0}$ for large $L$ should be a straight line with slope of order $1$.  

For each simulation run, $\kappa$ and the oasis density $n$ were input, $R_{\text{max}}$ and $L_{0}$ were calculated from (\ref{Rmax}) and (\ref{corr-fpp}), respectively, and a starting oasis was chosen near the center of the system.  The simulation then proceeded one infection event at a time, with infection times between oases generated using the distribution given in (\ref{FPTfortausimp}).  In order to speed up the simulation, we set a maximum distance $R_{\text{cut}}$ beyond which oases were effectively disconnected.  This allowed us to generate new oases "on-the-fly" as the simulation proceeded; together with our practice of throwing away information about an oasis once it was reached, this allowed us to only keep a small subset of oases in memory at any one time, thus allowing for the simulation of very large systems.  The value of $R_{\text{cut}}$ was chosen so as to make the probability of a missed event---that is, a jump event of size larger than $R_{\text{cut}}$ occurring over the course of the simulation---very small ($<10^{-3}$).

In early simulation runs, we found that our starting oasis would sometimes be isolated from the rest of the cluster, leading to larger-than-expected infection times with a large contribution from the time for the population to make the first jump.  In the limit as $L\to\infty$---the large-distance limit we're interested in---this contribution to the infection time, which does not grow with $L$, should become negligible, but for finite values of $L$ it can be important.  In order to eliminate this effect from our simulations without going to system sizes too large to be simulated in a reasonable amount of time, we allowed the population to ``find'' the cluster: we restarted the simulation once an oasis at least $2R_{\text{max}}$ from the starting oasis had been hit with the newly hit oasis as the new starting oasis.  The choice of $2R_{\text{max}}$ was admittedly arbitrary, but it did serve to eliminate the undesired effect from our simulations.


\begin{figure}
\includegraphics[scale=0.35]{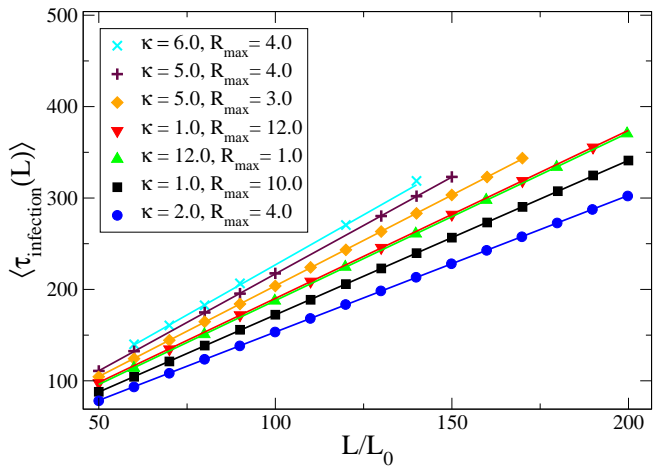}
\caption{\label{largesysFPT}First passage times across a large system shown for seven different combinations of $\kappa$ and $R_{\text{max}}$.  Error bars are not shown since they are, in most cases, smaller than the symbol size.  The lines represent best-fit lines for each $\kappa$, $R_{\text{max}}$.  The two lines with $\kappa R_{\text{max}}=12.0$ lie nearly on top of one another, as one would expect; we have omitted every other data point for each of these runs for clarity.  Note that the value of the slope (which is equal to $F(\kappa R_{\text{max}})$) increases as $\kappa R_{\text{max}}$ increases.}
\end{figure}

Once the population was restarted, the simulation continued one oasis infection event at a time.  When an oasis within a small distance $\delta\ll L_{0}$ of one of a set of concentric rings centered at the starting oasis was hit, the time and distance from the starting oasis were recorded; once all oases in some final ring were infected, the simulation ended.  The results of the simulation are shown in Fig.\ \ref{largesysFPT}.  The data confirms our picture of transport: the slopes of the best-fit lines through the data are indeed of order $1$, suggesting that $R_{\text{max}}$ is the correct length scale of the largest jumps the population must make on its way through the system and that $L_{0}$ is the correct length scale for the distance between these large jumps (of course, the population left behind the front edge will eventually make larger jumps to infect isolated oases, but this is unimportant in trying to determine the infection time).  Note that there are some ``missing'' points on the two lines with the highest $\kappa R_{\text{max}}$.  This is due to the presence of oases inside those rings which were not hit before the simulation time ended.  As $\kappa R_{\text{max}}$ is increased, such outlying oases take longer to hit, but since their ``extra'' contribution to the mean transit time does not scale with $L$, they do not affect our $L\to\infty$ results.

The slope for each line is equal to the scaling function $F(\kappa R_{\text{max}})$ for those values of $\kappa$ and $R_{\text{max}}$; note that $F(\kappa R_{\text{max}})$ appears to increase for increasing values of $\kappa R_{\text{max}}$.  This is likely due to that fact that, as $\kappa R_{\text{max}}$ increases, the correlation length $L_{0}$ increases, and thus the number of smaller oasis separations between the large oasis separations increases as well.  We do not understand this phenomenon completely, but it is seems a good candidate for further study; however, as the slopes are all of order $1$, an understanding of this phenomenon is hardly essential for making our present argument.  

\section{\label{concl}Conclusions, Remarks, and Future Work}
In this paper, we have examined transport in a reaction-diffusion system with disorder in the reaction rates.  Such systems have been used in the past to model bacterial population dynamics and the movement of plankton in the oceans.  Our model consists of particles which are allowed to diffuse with diffusion constant $D$ and compete for resources ($2A\to A$) everywhere with rate $b$, but which can only give birth ($A\to2A$) on small patches called oases at rate $y$ and which die ($A\to0$) everywhere else at rate $z$.  We have considered the limit in which the growth rate on the oases is very high and the oasis density is very low; in this limit, the time needed for a small population to grow on an oasis is much smaller than the typical time needed to jump from oasis to oasis, and thus transport can be thought of as a first passage process.  Because the population density traveling from one oasis to another is small, it is necessary to consider discreteness effects.  In order to determine the first passage time probability density function (FPT PDF) between two oases, we have employed a simplified model in which competition is ignored and the initially infected oasis is replaced by a particle source.  Simulations suggest that this model correctly predicts the FPT PDF for large oasis separations.

We have used an analogy with the theory of hopping conduction to argue that the largest oasis separations in the particle current-carrying cluster largely determine the time taken for a population to travel to a given target.  The scale of these separations can be found using continuum percolation theory, as in the hopping conduction problem.  There is a significant difference between the two problems: ours is dynamic, while the hopping conduction problem is not.  However, the use of results from first passage percolation theory suggest that the time scale for transit should still be determined by the largest oasis separations in the relevant particle current-carrying cluster.

There are certainly many future areas of study related to our work.  First off, there is the obvious question of what happens when the oases are not identical, but instead have their sizes and growth rates picked from some distribution.  One might hope that the theory of variable-range hopping \cite{shklovskii} would be useful in this case, though it remains to be seen whether the dynamic nature of the problem would make a fruitful mapping possible.  There is also the problem of determining the nature of the front that moves through a system like the one studied in this paper.  The velocity of such a front should be given roughly by $L_{0}/\langle T(R_{\text{max}},a)\rangle$, but its shape is an open question.  Finally, there is also the more general problem of RD wavefronts in media with quenched disorder, which is a challenge for future studies.

\begin{acknowledgments}
We would like to thank Bryan Clark, John Gergely, David Nelson, Mark Rudner, Nadav Shnerb, Richard Sowers, and Uwe T\"auber for helpful discussions.  This work was supported in part by NSF-DMR grants 03-14279 and 03-25939 (ITR) (UIUC Materials Computation Center), and by the L.S. Edelheit Family Biological Physics Fellowship.  We gratefully acknowledge the use of the Turing cluster maintained and operated by the Computational Science and Engineering Program at the University of Illinois.  Turing is a 1536-processor Apple G5 X-serve cluster devoted to high performance computing in engineering and science.  
\end{acknowledgments}

\appendix

\section{\label{1DMFsol}Solution of the Steady-State Mean-Field Equation in $d=1$}
The equation we need to solve is
\begin{equation}
0=D{\bar{c}_{ss}}^{\phantom{s}\prime\prime}(x)+U(x)\bar{c}_{ss}(x)-b\bar{c}_{ss}(x)^{2},
\end{equation}
where $U(x)=(y+z)\Theta(a-|x|)-z$ and the primes denote differentiation with respect to $x$.  To solve this, we can find solutions in the oasis ($|x|<a$) and desert ($|x|>a$) and then match at the boundaries.  In the desert, the relevant equation is $0=D{\bar{c}_{ss}}^{\phantom{s}\prime\prime}(x)-z\bar{c}_{ss}(x)-b\bar{c}_{ss}(x)^{2}$.  We define $u(c)\equiv u(\bar{c}_{ss}(x))\equiv{\bar{c}_{ss}}^{\phantom{s}\phantom{\prime}\prime}(x)$, which leads to the first order equation
\begin{equation}
0=u\frac{du}{dc}-\frac{z}{D}c-\frac{b}{D}c^2,
\end{equation}
where we have written $\bar{c}_{ss}(x)$ as $c$ for simplicity.  This equation can be integrated to give
\begin{equation}
u(c)=\frac{dc(x)}{dx}=-\sqrt{\frac{z}{D}}\,c(x)\sqrt{1+\frac{2bc(x)}{3z}},\label{udes}
\end{equation}
which can in turn be integrated to obtain the function $\bar{c}_{ss}(x)$ quoted on the second line of (\ref{c1Dssdes}).  A very similar procedure can be done for the area inside the oasis, leading to
\begin{equation}
u(c)=-\sqrt{\frac{2b}{3D}}\sqrt{c(x)^{3}-c(0)^{3}-\frac{3y}{2b}\left(c(x)^{2}-c(0)^{2}\right)},\label{uoas}
\end{equation}
which can also be integrated, leading to the function quoted on the first line of (\ref{c1Dssdes}).  Since the derivatives must match at the boundary, we can set (\ref{udes}) and (\ref{uoas}) equal at $|x|=a$ and obtain the following relation:
\begin{equation}
\bar{c}_{ss}(a)=\bar{c}_{ss}(0)\sqrt{\frac{3y-2b\bar{c}_{ss}(0)}{3(y+z)}}.
\end{equation}

\section{\label{yc}Derivation of the Formula for $y_{c}$}
The cutoff value of the growth rate $y$ below which a population placed on an oasis will die out as $t\to\infty$ can be estimated using the mean-field equation (\ref{FisherKPP}) with $b=0$.  For values of $y$ greater than the cutoff, the population will continue to increase without limit as $t\to\infty$; for $y<y_{c}$, the population will eventually die out.  At $y_{c}$, there will be a steady-state solution.  Hence, one way of finding the cutoff is to try to match solutions to the steady-state equation for $|\bm{x}|<a$ and $|\bm{x}|>a$ at $|\bm{x}|=a$.  Only along a certain line in parameter space will this be possible.

\begin{figure}
\includegraphics[scale=0.35]{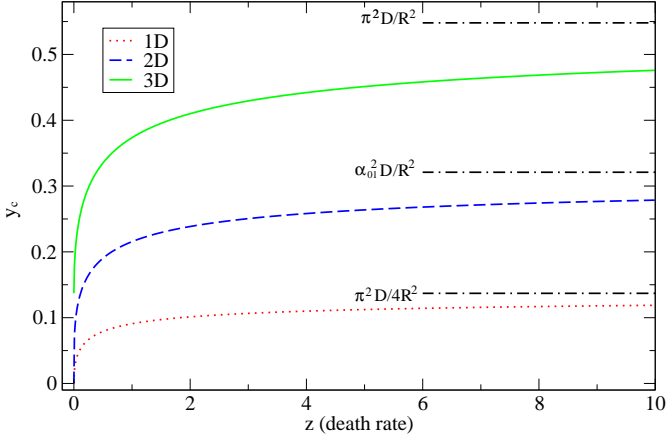}
\caption{\label{ycfigapp}Cutoff growth rate $y_{c}$ as a function of death rate $z$ with $a=3.0$, $D=0.5$.  Here $\alpha_{01}$ is the first zero of $J_{0}$.  Note that in one and two dimensions, an arbitrarily small growth rate with $z=0$ will allow a stable population to take hold; in three dimensions, $y_{c}(z=0)=\pi^{2}D/4a^{2}$.}
\end{figure}

In one dimension, the steady-state mean-field equation with $b=0$ is solved by $c(0)\cos(\sqrt{y/D}\,x)$ for $|x|<a$ and $c(a)e^{-\kappa (|x|-a)}$ for $|x|>a$.  Matching the functions and derivatives at $|x|=a$ leads to:
\begin{equation}
y_{c}=z\cot^{2}\left(\sqrt{\frac{y_{c}}{D}}\,a\right),\qquad(1\text{D})\label{ycutoff1Dapp}
\end{equation}
which is precisely (\ref{ycutoff1D}).  In two dimensions, a similar calculation leads to
\begin{equation}
y_{c}=z\left[\frac{J_{0}\left(\sqrt{\frac{y_{c}}{D}}\,a\right)K_{1}\left(\kappa a\right)}{J_{1}\left(\sqrt{\frac{y_{c}}{D}}\,a\right)K_{0}\left(\kappa a\right)}\right]^{2},\qquad(2\text{D})
\end{equation}
while in three dimensions we have obtained
\begin{equation}
y_{c}=z\tan^{2}\left(\pi-\sqrt{\frac{y_{c}}{D}}\,a\right).\qquad(3\text{D})
\end{equation}
These equations can be solved numerically to determine $y_{c}$.  A plot of $y_{c}$ as a function of $z$ in one, two, and three dimensions, with all other parameters fixed, is shown in Fig. (\ref{ycfigapp}).

\section{\label{asymp}Asymptotic Analysis of the Moments of $f_{N}(\bm{x},t)$}
In this appendix, we derive the results for the asymptotic moments of $f_{N}(\bm{x},t)$ quoted in Eqs. \ref{1Dcontmoms}, \ref{2Dcontmoms}, \ref{3Dcontmoms}, and \ref{1Dlattmoms}.  We start with the continuum case.  In any dimension, $P_{\text{none}}(R,a,t)\simeq\exp[-gY(R,a,t)]$, where $Y(R,a,t)$ is given by
\begin{equation}
Y(R,a,t)=\frac{(a/R)^{d/2-1}}{2\pi\imath}\int_{\mathcal{L}}ds\,\frac{e^{st}}{s^2}\,\frac{K_{d/2-1}\left(\sqrt{\frac{s+z}{D}}\,R\right)}{K_{d/2-1}\left(\sqrt{\frac{s+z}{D}}\,a\right)}.
\end{equation}
Although in $d=1$ and $d=3$ this Laplace transform machinery is unnecessary---we can simply perform the integral over time appearing in Eq. \ref{finalpnone}---it is easier to determine the asymptotic behavior of the moments of $f_{N}(R,a,t)$ in all dimensions by using these tools.  Changing variables to $p=s+z$ leads to $Y(R,a,t)=[(a/R)^{d/2-1}e^{-zt}/(2\pi\imath)]Q_{1}(R,a,t)$, where
\begin{equation}
Q_{1}(R,a,t)=\int_{\mathcal{L}}dp\,\frac{e^{pt}}{(p-z)^2}\,\frac{K_{d/2-1}\left(\sqrt{\frac{p}{D}}\,R\right)}{K_{d/2-1}\left(\sqrt{\frac{p}{D}}\,a\right)}.\label{Qint}
\end{equation}
\begin{figure}
\includegraphics[scale=0.45]{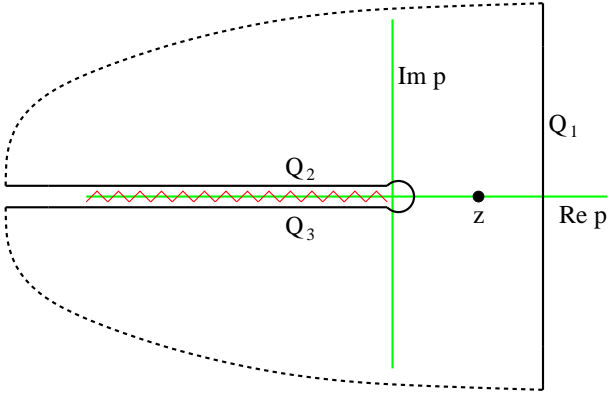}
\caption{\label{contourfig}Schematic of the contour integral which must be done to find $Y(R,a,t)$.  The dashed lines represent contributions to the integral that vanish as they are moved further from the origin.}
\end{figure}
This integral can be evaluated using contour integral techniques.  There is one second-order pole at $p=z$ and a branch cut which we will take to lie on the real $p$ axis from $p=0$ to $p=-\infty$.  Our countour will be taken to enclose the pole at $p=z$, and consists of three parts: $Q_{1}$, the value of which we wish to find; and $Q_{2}$ and $Q_{3}$, whose values must add with that of $Q_{1}$ to equal $2\pi\imath\Xi$, where $\Xi$ is the residue at $p=z$.  The space is shown schematically in Fig. \ref{contourfig}.  Using the residue theorem and changing integration variables to $u=-p$ gives:
\begin{eqnarray}
Q_{1}(R,a,t) & = & 2\pi\imath\,te^{zt}\,\frac{K_{\mu}\left(\kappa R\right)}{K_{\mu}\left(\kappa a\right)}\label{Qeqn}\\
& - & 2\pi\imath\frac{e^{zt}RK_{\mu+1}\left(\kappa R\right)K_{\mu}\left(\kappa a\right)}{\sqrt{4Dz}\left[K_{\mu}\left(\kappa a\right)\right]^{2}}\nonumber\\
& + & 2\pi\imath\frac{e^{zt}aK_{\mu}\left(\kappa R\right)K_{\mu+1}\left(\kappa a\right)}{\sqrt{4Dz}\left[K_{\mu}\left(\kappa a\right)\right]^{2}}\nonumber\\
& - & \int_{0}^{\infty}du\,\frac{e^{-tu}}{(u+z)^2}\,\frac{M_{\mu}(R,a,u)}{K_{\mu}\left(\imath\sqrt{\frac{u}{D}}\,a\right)K_{\mu}\left(-\imath\sqrt{\frac{u}{D}}\,a\right)},\nonumber
\end{eqnarray}
where we have used $\mu=d/2-1$ and $M_{\mu}(R,a,u)=2\imath\text{Im}\left[K_{\mu}\left(\imath\sqrt{\frac{u}{D}}\,R\right)K_{\mu}\left(-\imath\sqrt{\frac{u}{D}}\,a\right)\right]$.  We see that $Y(R,a,t)$ thus has the form $C_{1}t-C_{2}+C_{3}h(t)$, where the $C_{n}$ are constants in time and $h(t)$ is given by some complicated integral.  Since $Y(R,a,0)=0$, we can let $C_{3}=C_{2}$ and $h(0)=1$.  It should be clear that $h(\infty)=0$, and that $h(t)\leq1$ for all $t$.  This is enough to prove the asymptotic results for the moments of $f_{N}(R,a,t)$ quoted in Section \ref{preds}.  These moments are given by $\langle T^{j}(R,a)\rangle=j\int_{0}^{\infty}dt\,P_{\text{none}}(R,a,t)t^{j-1}$; plugging in the form for $Y(R,a,t)$ gives:
\begin{equation}
\langle T^{j}(R,a)\rangle=j\int_{0}^{\infty}dt\,e^{-g[C_{1}t-C_{2}(1-h(t))]}\,t^{j-1}
\end{equation}
The constant $C_{2}$ go to $0$ as $R\to\infty$, so one can Taylor expand $\exp[gC_{2}(1-h(t))]$ and arrive at
\begin{equation}
\langle T^{j}(R,a)\rangle=j\int_{0}^{\infty}dt\,e^{-gC_{1}t}\,t^{j-1}\left[1+gC_{2}(1-h(t))+\ldots\right]
\end{equation}
Keeping only the lowest order term, we get $\langle T^{j}(R,a)\rangle=j!{(gC_{1})}^{-j}$ as $R\to\infty$.  Looking at (\ref{Qeqn}), we see that $C_{1}=(a/R)^{d/2-1}K_{d/2-1}(\kappa R)/K_{d/2-1}(\kappa a)$.  We are now ready to plug in the functional forms for $K_{d/2-1}$ and arrive at the final asymptotic expressions for $\langle T^{j}(R,a)\rangle$:
\begin{eqnarray}
\langle T^{j}(x)\rangle &= & j!\,\frac{e^{\kappa|x|j}}{g^j}\qquad1\text{D}\\
\langle T^{j}(R,a)\rangle &= & j!\left[\frac{K_{0}(\kappa a)}{g\,K_{0}(\kappa R)}\right]^j\qquad2\text{D}\nonumber\\
\langle T^{j}(R,a)\rangle &= & j!\left(\frac{R}{a}\right)^j\,\frac{e^{\kappa(R-a)j}}{g^j}\qquad3\text{D},\nonumber
\end{eqnarray}
where $|x|=R-a$ (the distance from the origin to the edge of the oasis nearest the origin).  

\section{\label{convection}Convection Effects on First Passage Properties}
We wish to determine the effects of a small convection velocity on the first passage properties of a system.  Physically, such a convection velocity might represent the effects of a moving liquid medium in which the particles exist.  We start with a two-oasis system and use the linear model with a source to make analytical predictions possible.  To begin with, we replace the initially populated oasis with a source located at $\bm{R}$ and center our coordinate system in the middle of the target oasis of radius $a$.  The convection velocity $\bm{v}$ is taken to be constant in space.  In order to solve for $P_{\text{none}}(R,a,t)$, we must find $f_{1}(R,a,t)$, the single-particle FPT PDF.  This is done by solving for $p_{1}(\bm{x},t)$, the probability density function of a single particle released from the source at $\bm{R}$ at $t=0$, and then finding the probability flux into the oasis.

The diffusion equation governing $p_{1}(\bm{x},t)$, , is
\begin{equation}
\frac{\partial p_{1}(\bm{x},t)}{\partial t}=D\nabla^{2}p_{1}(\bm{x},t)-zp_{1}(\bm{x},t)-\bm{v}\cdot\nabla p_{1}(\bm{x},t).\label{diffwconvec}
\end{equation}
It is essential to simplify this equation before proceeding with a Laplace transform.  As with (\ref{1pdiff}), we can define a new function $\phi_{1}(\bm{x},t)=p_{1}(\bm{x},t)e^{zt}$ and eliminate the $-zp_{1}(\bm{x},t)$ term from the equation.  We can further define the function $\chi_{1}(\bm{x},t)$ via $\phi_{1}(\bm{x},t)=e^{\bm{v}\cdot\bm{x}/2D}\chi_{1}(\bm{x},t)$, leading to
\begin{equation}
\frac{\partial\chi_{1}(\bm{x},t)}{\partial t}=D\nabla^{2}\chi_{1}(\bm{x},t)-\frac{v^2}{4D}\,\chi_{1}(\bm{x},t).
\end{equation}
The last term on the right can be handled by defining $\chi_{1}(\bm{x},t)=\psi_{1}(\bm{x},t)e^{-v^2\,t/4D}$, leading to a simple diffusion equation for $\psi_{1}$.

The flux into the oasis can be used, as before, to find $f_{1}(R,a,t)$:
\begin{eqnarray}
f_{1}(R,a,t) & = & Da^{d-1}\int d\Omega\,\,\partial_{r}p_{1}(\bm{x},t)\\
 & = & Da^{d-1}e^{-(z+v^{2}/4D)t}\int d\Omega\,\,e^{\bm{v}\cdot\bm{x}/2D}\partial_{r}\psi_{1}(\bm{x},t),\nonumber
\end{eqnarray}
where $d\Omega$ is a differential element of angle in $2$D, and of solid angle in $3$D.  All that must be done is to find $\psi_{1}(\bm{x},t)$.  This function is the solution to a simple differential equation with initial condition $\psi_{1}(\bm{x},0)=e^{-\bm{v}\cdot\bm{R}/2D}\delta^{d}(\bm{x}-\bm{R})$, and is thus equal to $e^{-\bm{v}\cdot\bm{R}/2D}\phi_{1}(\bm{x},t)$, where $\phi_{1}(\bm{x},t)$ is the solution to the simple diffusion equation in the absence of convection.  Thus,
\begin{equation}
f_{1}(R,a,t)=Da^{d-1}e^{-(z+v^{2}/4D)t}e^{-\bm{v}\cdot\bm{R}/2D}\int d\Omega\,\,e^{\bm{v}\cdot\bm{x}/2D}\partial_{r}\phi_{1}(\bm{x},t).
\end{equation}
We are interested in the case where $R\gg a$, and so a decent approximation of $f_{1}(R,a,t)$ is given by
\begin{equation}
f_{1}(R,a,t)\simeq e^{-v^{2}t/4D}e^{\bm{v}\cdot\bm{R}/2D}f_{1}^{\bm{v}=0}(R,a,t).
\end{equation}
Note that in the above equation, we have reversed the sign of $\bm{R}$ since it is more natural to take the source as the origin rather than the center of the target oasis.  This result can be used to determine the moments of $f_{N}(R,a,t)$.  By making the replacements $z\to z+v^{2}/4D$ and $g\to ge^{\bm{v}\cdot\bm{R}/2D}$ in the expressions for the moments of $f_{N}(R,a,t)$, we arrive at the following expression, valid in any dimension:
\begin{equation}
\langle T^{j}(R,a)\rangle_{\bm{v}}=j!\left(\frac{R}{a}\right)^{(d-1)j/2}\frac{e^{-j\bm{v}\cdot\bm{R}/2D}e^{\kappa_{\bm{v}}(R-a)j}}{g^{j}},
\end{equation}
where $\kappa_{\bm{v}}=\sqrt{z/D+v^{2}/4D^{2}}$.


\begin{thebibliography}{44}
\bibitem[{not({\natexlab{a}})}]{noteonPoisson}
\bibinfo{note}{There is a subtle but ultimately unimportant inconsistency here:
  in writing the master equation for the linear model with a source, we
  implicitly assumed that the creation process at the source was a Poisson
  process. However, we are now treating it as a deterministic process in which
  particles are released at regular intervals. The mean-field equation for the
  linear model with a source is the same in either case, and so there is no
  ambiguity in fixing $g$.}
\bibitem[{not({\natexlab{b}})}]{noteonlengthscale}
\bibinfo{note}{We can relax this restriction slightly: the size of the largest
  link $R_{\text{max}}$ should be larger than $R_{\text{lin}}$, but the average
  distance between oases need not be.}
\end{thebibliography}
\end{document}